%% file: submit_v2.tex
\documentclass[amsmath,amssymb,twocolumn,prl,floatfix,showpacs,nofootinbib]{revtex4-1}
\usepackage{amsmath}
\usepackage{graphicx}
\usepackage{subfigure}
\usepackage{epstopdf}
\usepackage{color}
\usepackage{multirow}
\usepackage{setspace}
\usepackage{overpic}
\usepackage{amssymb}
\usepackage[colorlinks,urlcolor=blue, citecolor=blue,linkcolor=blue] {hyperref}
\usepackage{lineno}
\usepackage{bm}
\usepackage{rotating}
\usepackage[utf8]{inputenc}
\let\oldequation\equation
\let\oldendequation\endequation
\renewenvironment{equation}
 {\linenomathNonumbers\oldequation}
 {\oldendequation\endlinenomath}

\begin{document}

\title{\bf \boldmath
Observation of the Doubly Cabibbo-Suppressed Decay $D^+\to K^+\pi^+\pi^-\pi^0$ and Evidence for $D^+\to K^+\omega$
}

\input{BESIII_authors}

\begin{abstract}
Using $2.93\,\rm fb^{-1}$ of $e^+e^-$ collision data collected
at a center-of-mass energy of 3.773\,GeV with the BESIII detector, the first observation of the doubly Cabibbo-suppressed decay $D^+\to K^+\pi^+\pi^-\pi^0$ is reported. After removing decays that contain narrow intermediate resonances, including $D^+\to K^+\eta$, $D^+\to K^+\omega$, and $D^+\to K^+\phi$, the branching fraction of the decay $D^+\to K^+\pi^+\pi^-\pi^0$ is measured to be $(1.13 \pm 0.08_{\rm stat} \pm 0.03_{\rm syst})\times 10^{-3}$.
The ratio of branching fractions of $D^+\to K^+\pi^+\pi^-\pi^0$ over $D^+\to K^-\pi^+\pi^+\pi^0$ is found to be $(1.81\pm0.15)$\%, which corresponds to $(6.28\pm0.52)\tan^4\theta_C$, where $\theta_C$ is the Cabibbo mixing angle. This ratio is significantly larger than the corresponding ratios for other doubly Cabibbo-suppressed decays. The asymmetry of the branching fractions of charge-conjugated decays $D^\pm\to K^\pm\pi^\pm\pi^\mp\pi^0$ is also determined, and no evidence for $CP$ violation is found.
In addition, the first evidence for the $D^+\to K^+\omega$ decay, with a statistical significance of 3.3$\sigma$, is presented and the branching fraction is measured to be ${\mathcal B}(D^+ \to K^+ \omega)=({5.7^{+2.5}_{-2.1}}_{\rm stat}\pm0.2_{\rm syst})\times10^{-5}$.

\end{abstract}


\maketitle

\footnotetext{Corresponding authors: \\
xpan19@fudan.edu.cn, luot@fudan.edu.cn, mahl@ihep.ac.cn}

\oddsidemargin  -0.2cm
\evensidemargin -0.2cm

Doubly Cabibbo-suppressed (DCS) decays of $D$ mesons can provide unique insight into weak decay mechanisms of charmed hadrons. To date, DCS decays of charmed hadrons remain relatively unexplored~\cite{pdg2020}.
The naive expectation
for the DCS decay rate relative to its Cabibbo-favored (CF) counterpart
\cite{Lipkin,theory_1}
is of the order ${\rm tan}^4\theta_C$\,$\sim$\,0.29\%, where $\theta_C$
is the Cabibbo mixing angle.
The known ratios of DCS and CF decay rates~~\cite{pap3} roughly support
this expectation, with  the exception of $D^+\to K^-\pi^+\pi^+$ \cite{pap201}
where the ratio is doubled due to identical particles
in the final state.
A measurement of the branching fraction (BF) of $D^+\to K^+\pi^+\pi^-\pi^0$
and a comparison with its CF counterpart provides a  crucial test of this expectation.

In theory the BFs of $D\to VP$ decays, where $V$ and $P$ denote
vector and pseudoscalar mesons, respectively, can be calculated
after incorporating quark SU(3)-flavor symmetry and symmetry breaking
as well as charge-parity ($CP$)
violation~\cite{ref5,theory_a,theory_5,theory_4,theory_3,theory_2,theory_1,zzxing,ref6}.
The experimental information on DCS $D\to VP$ decays is
currently limited.
Investigation of $D^+\to K^+\pi^+\pi^-\pi^0$ offers an ideal
opportunity to determine the BF of $D^+\to K^+\omega$ with $\omega \to\pi^+\pi^-\pi^0$, where $\omega$ stands for $\omega(782)$ throughout
this Letter.
The result is important to improve our understanding
of quark SU(3)-flavor symmetry and symmetry breaking
and also benefits theoretical calculations of $CP$ violation~\cite{ref5,theory_a,theory_5,theory_4,theory_3,theory_2,theory_1,zzxing,ref6}.

In the Standard Model, $CP$ violation in the weak decays of hadrons
arises due to a single irreducible phase in the
Cabibbo-Kobayashi-Maskawa matrix~\cite{pap1}.
$CP$ violation in charmed-hadron decays is expected to be small, up to a few $10^{-3}$ for singly Cabibbo-suppressed processes,
and much smaller for CF and DCS processes~\cite{ref5,pap02}.
In the past two decades, $CP$ violation in charm sector has been extensively explored~\cite{pap05}. In 2019, the LHCb collaboration reported an observation of $CP$ violation in the singly-Cabibbo-suppressed decays $D^0\to K^+K^-$ and $D^0\to \pi^+\pi^-$~\cite{lhcb_D_CP}.
Searching for $CP$ violation in DCS decays offers complementary information about $CP$ violation in the charm sector.

This Letter reports the first measurement of the absolute BFs of the DCS decays $D^+\to K^+\pi^+\pi^-\pi^0$ and $D^+\to K^+\omega$.
Charge-conjugated decays are always implied unless stated otherwise.
The $CP$ asymmetry of $D^\pm \to K^\pm\pi^\pm\pi^\mp\pi^0$ is also presented.

This work is performed by using 2.93\,fb$^{-1}$ of $e^+e^-$ collision data~\cite{lum_bes3} collected with the BESIII detector
at the center-of-mass energy of $\sqrt s=$ 3.773~GeV.
This energy is
near the resonance peak of the $\psi(3770)$, which predominantly decays
into $D\bar D$ ($D$ denotes $D^0$ or $D^+$) pair.
The two $D$ mesons are produced close to rest in the centre of mass frame without accompanying hadron(s),
thereby offering ideal environment to study $D$ meson decays with
the double-tag (DT) technique, pioneered by the Mark III Collaboration~\cite{mark3}.

Details about the design and performance of the BESIII detector are given in Refs.~\cite{BESIII,whitepaper}.
Simulated samples produced with a {\sc
Geant4}-based~\cite{geant4} Monte Carlo (MC) package, which
includes the geometric description of the BESIII detector and the
detector response, are used to determine the detection efficiency
and to estimate backgrounds. The simulation includes the beam
energy spread and initial state radiation (ISR) in the $e^+e^-$
annihilations modeled with the generator {\sc
kkmc}~\cite{kkmc}.
The signal of $D^+\to K^+\pi^+\pi^-\pi^0$ is simulated using an MC generator that incorporates
the resonant decays $D^+\to K^*(892)^0\rho(770)^+$, $K^*(892)^+\rho(770)^0$,
$K^+\eta$, $K^+\omega$, the phase space decay $D^+\to K^+\pi^+\pi^-\pi^0$,
and possible interferences. The parameters of the generator
have been tuned to reach a good data-MC agreement in distributions
of the daughter particle momenta
and the invariant masses of each two- and three-body particle combinations.
The signal of $D^+\to K^+\omega$ is simulated using an MC generator which simulates
pseudoscalar meson decays into vector meson and scalar meson~\cite{evtgen}.
The background is studied using an inclusive MC sample that consists of the
production of $D\bar{D}$
pairs with consideration of quantum coherence for all neutral $D$
modes, the non-$D\bar{D}$ decays of the $\psi(3770)$, the ISR
production of the $J/\psi$ and $\psi(3686)$ states, and the
continuum processes incorporated in {\sc kkmc}.
The known decay modes are modeled with {\sc
evtgen}~\cite{evtgen} using the known BFs taken from the
Particle Data Group~\cite{pdg2020}, while the remaining unknown decays
from the charmonium states are modeled with {\sc
lundcharm}~\cite{lundcharm}. Final state radiation
from charged final state particles is incorporated with the {\sc
photos} package~\cite{photos}.

We obtain the BFs by reconstructing signal $D^+$ decays in events with $D^-$ decays reconstructed in one of the three decay modes $D^-\to K^+\pi^-\pi^-$, $D^-\to K^0_S\pi^-$, and $D^-\to K^+\pi^-\pi^-\pi^0$. If a $D^-$ meson is found, it is referred to as a single-tag (ST) candidate. An event in which a signal $D^+$ decay and an ST $D^-$ are simultaneously found is referred as a double-tag event.
The BF of the signal decay is given by
\begin{equation}
\label{eq:br}
{\mathcal B}_{{\rm sig}} = \frac{N_{\rm DT}}{\sum\limits_{i=1}^{3}{N^i_{\rm ST}(\epsilon_{\rm DT}^i/\epsilon_{\rm ST}^i)}},
\end{equation}
where ${ N_{\rm DT}}$ is the number of events with any $D^-$ tag and a signal candidate, $\epsilon_{\rm DT}^i$ is the signal selection efficiency for an event with a $D^-$ in the $i$-th tag mode, and $N_{\rm ST}^i$ and $\epsilon_{\rm ST}^i$ are the number of tags and reconstruction efficiency for $D^-$ candidates in mode $i$.

\begin{figure}[htp]
  \centering
\includegraphics[width=1\linewidth]{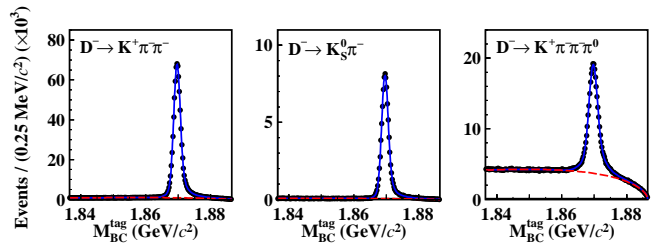}
  \caption{\small
Fits to the $M_{\rm BC}$ distributions of the ST $D^-$ candidates. Data are shown as dots with error bars. The blue solid and red dashed curves are the fit results and the fitted backgrounds, respectively.}
\label{fig:STfit}
\end{figure}

The $K^0_S$ and $\pi^0$ candidates are reconstructed via $K^0_S\to\pi^+\pi^-$ and $\pi^0\to\gamma\gamma$, respectively.
For the reconstruction and identification of $K^\pm$, $\pi^\pm$,
$K^0_S$, and $\pi^0$ we use the same
criteria as in
Refs.~\cite{epjc76,cpc40,bes3-pimuv,bes3-Dp-K1ev,bes3-etaetapi,omegamunu,papnew,papnew1,kkpipi,b1enu}.
The tagged $D^-$ mesons are selected using two variables, the energy difference
\begin{equation}\label{def_delE}
\Delta E_{\rm tag} \equiv E_{D^-} - E_{\rm b},
\end{equation}
and the beam-constrained mass
\begin{equation}\label{def_mbc}
M_{\rm BC}^{\rm tag} \equiv \sqrt{E^{2}_{\rm b}-|\vec{p}_{D^-}|^{2}},
\end{equation}
where $E_{\rm b}$ is the beam energy, and $\vec{p}_{D^-}$ and $E_{D^-}$ are the momentum and the energy of the $D^-$ candidate in the $e^+e^-$ rest frame. For each tag mode, if there are multiple combinations, the one giving the minimum $|\Delta E_{\rm tag}|$ is retained for further analysis.
The tagged $D^-$ are required to satisfy
$\Delta E_{\rm tag}\in(-55,\, 40)$\,MeV for the decay mode containing a $\pi^0$,
and $\Delta E_{\rm tag}\in(-25,\, 25)$\,MeV for the other decay modes.
The yields of ST $D^-$ mesons were obtained from maximum likelihood fits to the $M_{\rm BC}^{\rm tag}$
distributions of the accepted ST candidates~\cite{epjc76,cpc40,bes3-pimuv,bes3-Dp-K1ev,bes3-etaetapi,omegamunu}.
The fit results are shown in Fig.~\ref{fig:STfit}.
The total ST $D^-$ yield is ${N_{{\rm ST}}=}(1150.3\pm1.5_{\rm stat})\times 10^3$.

\begin{figure}[tp]
  \centering
\includegraphics[width=1\linewidth]{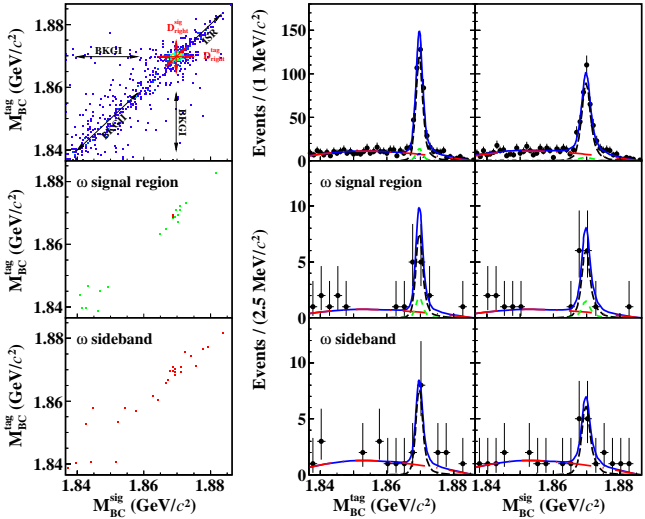}
  \caption{\small
Distributions of (left column) $M_{\rm BC}^{\rm tag}$ vs. $M_{\rm BC}^{\rm sig}$, and the projections of the corresponding 2D fits on (middle column) $M_{\rm BC}^{\rm tag}$ and (right column) $M_{\rm BC}^{\rm sig}$, for the DT candidate events of $D^-\to$ all tags vs. $D^+\to K^+\pi^+\pi^-\pi^0$. The top, middle, and bottom rows correspond to all events, events lying in $\omega$ signal region, and those falling in $\omega$ sideband region, respectively. In the figures of the middle and right columns, data are shown as dots with error bars; the blue solid, black dashed, blue dot-dashed,
red dot-long-dashed, and green dashed curves
denote the overall fit results, signal, BKGI, BKGII,
and peaking background components, respectively.}
\label{fig:2Dfit}
\end{figure}

The signal $D^+$ candidates are reconstructed from the particles
that have not been used for the tagged $D^-$ reconstruction.
They are identified using the energy difference
and the beam-constrained mass of the signal side,
$\Delta E_{\rm sig}$ and $M_{\rm BC}^{\rm sig}$,
calculated similarly to Eqs.~(\ref{def_delE}) and~(\ref{def_mbc}), respectively, with $D^-$ replaced by $D^+$.
If there are multiple combinations, caused mainly due to incorrectly $\pi^0$, the one giving the minimum $|\Delta E_{\rm sig}|$ is retained for further analysis.
The signal side is required to be within $\Delta E_{\rm sig}\in (-58,45)\,{\rm MeV}$.
The invariant mass of the $\pi^+\pi^-$ pair
must satisfy the condition $|M_{\pi^+\pi^-}-M_{K_S^0}|\,>\,20$\,MeV/$c^2$ to reject the dominant peaking background from the singly Cabibbo-suppressed decay $D^+\to K_S^0 K^+\pi^0$. This requirement corresponds to about $\pm 5\sigma$ of the experimental resolution.
To suppress non-$D^+D^-$ events,
the opening angle between the $D^+$ and $D^-$ candidates is required to be greater than $160^\circ$,
which results in a loss of 6\% of the signal but rejects 34\% of the background contributions.
The top-left panel of Fig.~\ref{fig:2Dfit} shows the $M_{\rm BC}^{\rm tag}$ vs. $M_{\rm BC}^{\rm sig}$ distribution of the accepted candidates for $D^+\to K^+\pi^+\pi^-\pi^0$ in data. The comparison of two-body and three-body mass distributions of the accepted $D^+\to K^+\pi^+\pi^-\pi^0$ candidate events
can be found in the supplemental material~\cite{supplemental}.

\begin{figure}[htp]
  \centering
\includegraphics[width=1.0\linewidth]{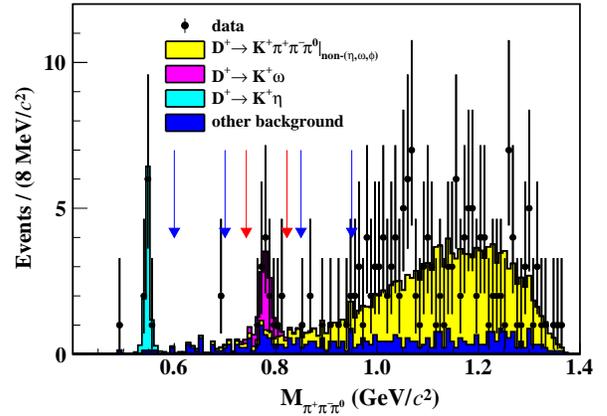}
  \caption{\small
  Distribution of $M_{\pi^+\pi^-\pi^0}$ for $D^+\to K^+\pi^+\pi^-\pi^0$ candidates in data (dots with error bars). Histograms in yellow, pink, and cyan are the signal MC events of $D^+\to K^+\pi^+\pi^-\pi^0|_{\rm non{\text -}\eta,\omega,\phi}$, $D^+\to K^+\omega$, and $D^+\to K^+\eta$ normalized with individual BFs and efficiencies, and blue histogram is the background estimated using the inclusive MC sample, scaled to the rest event yield in data. Events have been selected using $M_{\rm BC}^{\rm tag\, (sig)}\in(1.863,1.875)$~GeV/$c^2$ and all other requirements for $D^+\to K^+\omega$
except for the $\omega$ signal mass window. The red arrows denote the $\omega$ signal region. The blue arrows denote the $\omega$ sideband regions.}
\label{fig:m3pi}
\end{figure}

Furthermore, the $D^+\to K^+\omega$ candidates are selected from events with $\pi^+\pi^-\pi^0$ invariant mass within $|M_{\pi^+\pi^-\pi^0}-M_{\omega}|<40$~MeV/$c^2$,
where $M_{\omega}$ is the nominal mass of the $\omega$ meson~\cite{pdg2020}. This requirement is set by taking into account both the natural width of the $\omega$ meson and the invariant mass resolution. To suppress non-$\omega$ backgrounds, the $\omega$ helicity angle is required to satisfy $|\cos\theta_\omega|\,>\,0.57$, where $\theta_\omega$ is the opening angle between the normal to the $\omega\to\pi^+\pi^-\pi^0$ decay plane and the direction of the $D^+$ meson in the $\omega$ rest frame. Moreover, the normalized slope parameter $\lambda/\lambda_{\rm max}$, introduced in Ref.~\cite{pap25}, is required to be greater than 0.21, where the criterion is based on an optimization using the inclusive MC sample. The middle-left and bottom-left figures of Fig.~\ref{fig:2Dfit} show the $M_{\rm BC}^{\rm tag}$ vs. $M_{\rm BC}^{\rm sig}$ distributions of the accepted candidates with the aforementioned additional requirements for $D^+\to K^+\pi^+\pi^-\pi^0$ in data, with $M_{\pi^+\pi^-\pi^0}$ in the $\omega$ signal region and the $\omega$ sideband region, defined as $M_{\pi^+\pi^-\pi^0}\in (0.60,0.70)\cup(0.85,0.95)$ GeV/$c^2$, respectively.
Figure~\ref{fig:m3pi} shows the definitions of the $\omega$ signal and sideband regions.

In the $M_{\rm BC}^{\rm tag}$ vs. $M_{\rm BC}^{\rm sig}$ distributions, as shown in the left column of Fig.~\ref{fig:2Dfit},
signal events concentrate around $M_{\rm BC}^{\rm tag} = M_{\rm BC}^{\rm sig} = M_{D}$,
where $M_{D}$ is the nominal mass of the $D^+$ meson~\cite{pdg2020}.
Background events are divided into three categories.
The first (BKGI) is from events with correctly reconstructed $D^+$ ($D^-$) and incorrectly
reconstructed $D^-$ ($D^+$). This background is distributed along
the horizontal and vertical bands.
The second (BKGII) describes events found along the diagonal,
which are mainly from the $e^+e^- \to q\bar q$ processes.
The third (BKGIII) consists of uniformly distributed
events in which both the tagged $D^-$ and the signal $D^+$
are reconstructed incorrectly.
For the decay $D^+\to K^+\pi^+\pi^-\pi^0$ the peaking backgrounds
from $D^+\to K^+K^-(\to\pi^-\pi^0)\pi^+$ decays and
from the residual $D^+\to K^0_S(\to \pi^+\pi^-)K^+\pi^0$ events are evaluated
using the MC simulations. For the decay $D^+\to K^+\omega$,
the peaking background contributions are dominated by the
non-$\omega$ decays $D^+\to K^+\pi^+\pi^-\pi^0$.
This peaking background has the same event topology as the signal
and is estimated using data events in the $\omega$ sideband region defined above.

To extract the DT yields, a two-dimensional (2D) unbinned maximum likelihood fit
is performed on the corresponding $M_{\rm BC}^{\rm tag}$ vs. $M_{\rm BC}^{\rm sig}$ distribution.
The 2D probability density function (PDF) for the signal is taken from the MC simulation. The PDFs of background contributions are constructed as~\cite{bes3-2D,cleo-2D,bes3-etaetapi,papnew1,kkpipi}
\begin{itemize}
\item BKGI: $b(x)\cdot c_y(y;E_{\rm b},\xi_{y}) + b(y)\cdot c_x(x;E_{\rm b},\xi_{x})$,
\item BKGII: $c_z(z;\sqrt{2}E_{\rm b},\xi_{z}) \cdot g(k;0,\sigma_k)$,
\item BKGIII: $c_x(x;E_{\rm b},\xi_{x}) \cdot c_y(y;E_{\rm b},\xi_{y})$.
\end{itemize}
Here, $x=M_{\rm BC}^{\rm tag}$, $y=M_{\rm BC}^{\rm sig}$, $z=(x+y)/\sqrt{2}$, and $k=(x-y)/\sqrt{2}$.
The functions $b(x)$ and $b(y)$ are the one-dimensional signal shapes taken from the MC simulation.
The function $c_f$ is the ARGUS function~\cite{ARGUS} defined as
\begin{equation}
c_f\left(f; E_{\rm b}, \xi_f\right) = A_f f (1 - \frac {f^2}{E_{\rm b}^2})^{\frac{1}{2}} e^{\xi_f (1-\frac {f^2}{E_{\rm b}^2})},
\end{equation}
where $f$ denotes $x$, $y$, or $z$, $E_{\rm b}$ is fixed at 1.8865 GeV, $A_f$ is a normalization factor,
and $\xi_f$ is a fit parameter.
The function $g(k;\sigma_k)$ is a Gaussian distribution with a mean of zero and a standard deviation $\sigma_k=\sigma_0 \cdot(\sqrt{2}E_{\rm b}-z)^p$,
where $\sigma_0$ and $p$ are parameters determined by the fit.
For the decay $D^+\to K^+\pi^+\pi^-\pi^0$ the yields and shapes of the peaking background contributions are fixed
to the expectation from the MC simulations.
The BKGIII component is ignored due to limited data.
All other parameters are left free.

To extract the signal yield of $D^+\to K^+\omega$,
simultaneous 2D fits are performed on the events
in the $\omega$ signal and sideband regions.
The background PDFs are fixed to the shapes obtained
from the $D^+\to K^+\pi^+\pi^-\pi^0$ fit.
The ratio of the background
yield in the $\omega$ sideband region and
in the $\omega$ signal region is fixed to the value
$f_{\omega}=4.12\pm0.08$ obtained using
the $D^+\to K^+\pi^+\pi^-\pi^0$ MC simulation.
The reliability of the choice and normalization of the nominal $\omega$ sideband region has been further verified
by using those events with $M_{\pi^+\pi^-\pi^0}\in (0.85,1.35)$ GeV/$c^2$ arbitrarily.
Both BKGI and BKGIII components are ignored in these two fits because of limited data.

The spectra in the middle and right columns in Fig.~\ref{fig:2Dfit} show
the projections on $M_{\rm BC}^{\rm tag}$ and $M_{\rm BC}^{\rm sig}$ of the 2D fits to data.
For both signal decay modes the statistical significance is evaluated as $\sqrt{-2{\rm ln}(\mathcal L_0/\mathcal L_{\rm max})} $,
where $\mathcal L_{\rm max}$ is the maximum likelihood of the nominal fit and $\mathcal L_0$ is the likelihood of the fit excluding the signal PDF, and the degree of freedom is assumed to be 1.
The statistical significance is found to be $23.3\sigma$ for $D^+\to K^+\pi^+\pi^-\pi^0$ and $3.3\sigma$ for $D^+\to K^+\omega$.
For $D^+\to K^+\omega$, the effect of the fluctuation of the $\omega$ sideband events has been considered in the simultaneous fit.

The numbers of ${ N_{{\rm DT}}}$ and $\epsilon^{}_{{\rm sig}}$
as well as the obtained BFs of the two decays
are summarized in the first two rows of Table~\ref{tab:DT}.

\begin{table*}[htp]
\centering
\caption{\label{tab:DT}
\small
The ST and DT yields in data ($N_{\rm ST}$ and $N_{\rm DT}$), the signal efficiencies ($\epsilon_{\rm sig}$), and the obtained BFs before~($\mathcal B_{\rm sig}$) and after~(${\mathcal B}^*_{\rm sig}$) removing the contribution from $D^+\to K^+\eta$, $K^+\omega$, and $K^+\phi$~\cite{note_BF}. Here, we ignore the possible interferences between these two-body decays and the other processes in $D^+\to K^+\pi^+\pi^-\pi^0$. The uncertainties are statistical only. }
\begin{tabular}{lccccc}
\hline\hline
Decay mode  &$N_{\rm ST}~(\times\,10^{3})$          & $N_{\rm DT}$ & $\epsilon_{\rm sig}$\,(\%)& $\mathcal B_{\rm sig}~(\times\,10^{-3})$&$\mathcal B_{\rm sig}^{*}~(\times\,10^{-3})$\\ \hline
$D^\pm\to K^\pm\pi^\pm\pi^\mp\pi^0$ &$1150.3\pm1.5$&$350\pm22$&$25.03\pm0.13$&$1.21\pm0.08$&$1.13\pm0.08$ \\

$D^\pm\to K^\pm\omega$
&$1150.3\pm1.5$&$9.2^{+4.0}_{-3.4}$&$14.14\pm0.09$&$(5.7^{+2.5}_{-2.1})\,\times 10^{-2}$&- \\ \hline
$D^+\to K^+\pi^+\pi^-\pi^0$ &$573.5\pm1.0$&$181\pm15$&$25.20\pm0.18$&$1.25\pm0.11$&$1.17\pm0.11$\\
$D^-\to K^-\pi^-\pi^+\pi^0$ &$572.7\pm1.0$&$165\pm15$&$24.95\pm0.18$&$1.16\pm0.11$&$1.08\pm0.11$\\

\hline\hline
\end{tabular}
\end{table*}

With the DT method, most of the uncertainties related
to the ST selection are negligible.
The systematic uncertainties arise from the following
sources and are estimated relative to the measured BFs.
The uncertainty on the total ST $D^-$ yield
is due to the fit to the $M_{\rm BC}^{\rm tag}$ distributions
and is estimated to be
0.5\%~\cite{epjc76,cpc40,bes3-pimuv}. The tracking and PID efficiencies of $K^\pm$ and $\pi^\pm$ are studied with DT $D\bar D$ hadronic events. A small difference between the $K^\pm$ tracking efficiency in data and in MC simulation is found, but those for the efficiencies of $K^\pm$ PID, $\pi^\pm$ tracking and $\pi^\pm$ PID are negligible. The averaged data$\text{-}$MC difference of $K^\pm$ tracking efficiency weighted by the momentum spectrum of signal MC events is $1.8\%$. After correcting the MC efficiencies by this averaged data$\text{–}$MC difference, the systematic uncertainties of tracking efficiencies are estimated to be 0.3\% per $K^\pm$ or $\pi^\pm$. The systematic uncertainties originating from PID efficiencies are assigned as 0.3\% per $K^\pm$ or $\pi^\pm$.
The efficiency of reconstructing a $\pi^0$ meson is investigated by using the DT $D\bar D$ hadronic decay samples of $D^0\to K^-\pi^+$, $K^-\pi^+\pi^+\pi^-$ vs. $\bar D^0\to K^+\pi^-\pi^0$, $K^0_S\pi^0$~\cite{epjc76,cpc40}. The averaged data-MC difference of the $\pi^0$ reconstruction efficiencies, weighted by the momentum spectra of signal MC events, is $0.7\%$ per $\pi^0$. After correcting the MC efficiencies by this averaged data$\text{–}$MC difference
the systematic uncertainty arising from $\pi^0$ reconstruction is estimated as 0.8\% per $\pi^0$.
The uncertainties of the quoted BFs of $\omega\to\pi^+\pi^-\pi^0$ and $\pi^0\to\gamma\gamma$ decays are 0.8\% and 0.03\%~\cite{pdg2020}, respectively.

To estimate the systematic uncertainty from the 2D fit,
the measurements are repeated
by varying the signal shape, the endpoint of the ARGUS function, and the fixed number of peaking background events (by varying $\pm 1\sigma$ of the quoted BFs of the dominant peaking backgrounds of $D^+\to K^0_S(\to \pi^+\pi^-)K^+\pi^0$ and $D^+\to K^+K^-(\to\pi^-\pi^0)\pi^+$). Quadratically summing over the changes of the BFs gives the systematic uncertainties, which are 0.9\% for $D^+\to K^+\pi^+\pi^-\pi^0$ and negligible for $D^+\to K^+\omega$.
The systematic uncertainty related to the $D^+D^-$ opening angle requirement
is assigned as 0.5\% based on DT events where
the signal decays are replaced by the CF $D^+\to K^-\pi^+\pi^+\pi^0$ channel.
The systematic uncertainty associated with the $\Delta E_{\rm sig}$
requirement is evaluated to be 0.2\%, estimated by
smearing the $\Delta E_{\rm sig}$ distribution for signal MC events.
The systematic uncertainty due to $K^0_S$ rejection is negligible
since the mass resolution is well reproduced by the MC simulation.
The boundaries of the $\omega$ sideband regions
were varied by $\pm5$ MeV/$c^2$ and the corresponding
uncertainty was found to be negligible.
The limited number of simulated events contributes 0.5\% uncertainty
for $D^+\to K^+\pi^+\pi^-\pi^0$ and 0.6\% for $D^+\to K^+\omega$.
The systematic uncertainty related to the MC modeling
for $D^+\to K^+\pi^+\pi^-\pi^0$ is assigned to be 1.3\%, which is the difference of the DT efficiencies with and without involving the less significant decays of $D^+\to K^+\eta$, $K^+\omega$, and $K^+\phi$, and the effects of high excited states are negligible.
For $D^+\to K^+\omega$, the systematic uncertainties of the MC modeling are mainly from the imperfect simulations on
$\cos\theta_\omega$ and $\lambda/\lambda_{\rm max}$.
They are estimated using the DT events $D^0\to K^0_S\omega$ vs. $\bar D^0\to K^+\pi^-$, $K^+\pi^-\pi^0$, and $K^+\pi^-\pi^-\pi^+$.
The differences of the acceptance efficiencies of the $\cos\theta_\omega$ and $\lambda/\lambda_{\rm max}$ requirements between data and MC simulations,
3.0\% and 1.2\%, are assigned as the corresponding systematic uncertainties, respectively.
The uncertainty on the scale factor
$f^{\rm sid/sig}_{\omega}$ results in 0.6\% uncertainty
on the $D^+\to K^+\omega$ signal.

The total systematic uncertainty of the BF measurement is 2.3\% for $D^+\to K^+\pi^+\pi^-\pi^0$ and 3.8\% for $D^+\to K^+\omega$, obtained by adding the above effects quadratically.

The BFs of the charge-conjugated decays $D^+\to K^+\pi^+\pi^-\pi^0$ and $D^-\to K^-\pi^-\pi^+\pi^0$, ${\mathcal B}_{D^+\to K^+\pi^+\pi^-\pi^0}$ and ${\mathcal B}_{D^-\to K^-\pi^-\pi^+\pi^0}$, are measured separately.
The asymmetry of these two BFs is determined as
\begin{equation}
{{\mathcal A}_{CP}^{D^\pm\to K^\pm\pi^\pm\pi^\mp\pi^0}}=\frac{{\mathcal B}_{D^+\to K^+\pi^+\pi^-\pi^0}-{\mathcal B}_{D^-\to K^-\pi^-\pi^+\pi^0}}{{\mathcal B}_{D^+\to K^+\pi^+\pi^-\pi^0}+{\mathcal B}_{D^-\to K^-\pi^-\pi^+\pi^0}}.
\end{equation}
The corresponding ST yields, DT yields, signal efficiencies, and the obtained BFs are summarized in the last two rows of Table~\ref{tab:DT}.
The asymmetry is determined to be ${\mathcal A}_{CP}^{D^\pm \to K^\pm\pi^\pm\pi^\mp\pi^0} = (-0.04\pm0.06_{\rm stat}\pm0.01_{\rm syst})$,
where the systematic uncertainties of tracking and PID of the $\pi^+\pi^-$ pair, $\pi^0$ reconstruction, quoted BFs, and MC modeling cancel.
Other systematic uncertainties are estimated separately as above.
No evidence for $CP$ violation is found.

In summary, using $2.93\,\rm fb^{-1}$ of data taken at $\sqrt{s}=3.773$\,GeV with the BESIII detector, the first observation and BF measurement of the DCS decay $D^+\to K^+\pi^+\pi^-\pi^0$ are presented. Removing the contribution of the known decays $D^+\to K^+\eta$, $K^+\omega$, and $K^+\phi$~\cite{note_BF} and ignoring the possible interferences between these decays and the other processes in $D^+\to K^+\pi^+\pi^-\pi^0$, we obtain ${\mathcal B}^*_{D^+\to K^+\pi^+\pi^-\pi^0}=(1.13 \pm 0.08_{\rm stat} \pm 0.03_{\rm syst})\times 10^{-3}$, which is the largest among all known DCS decays in the charm sector.
The evidence for the decay $D^+\to K^+\omega$ is found and its BF is measured to be $({5.7^{+2.5}_{-2.1}}_{\rm stat}\pm0.2_{\rm syst})\times10^{-5}$.
This BF is consistent with theoretical predictions that incorporate quark SU(3)-flavor symmetry and symmetry breaking~\cite{theory_2},
but disfavors predictions based on quark SU(3)-flavor symmetry without symmetry breaking~\cite{theory_3,theory_1}
and predictions based on the pole model~\cite{yufs} by 1.8$\textendash$2.8$\sigma$.
This result will benefit future calculations of $CP$ violation in the charm sector~\cite{zzxing,theory_a,theory_1,theory_2,theory_3,theory_4,theory_5,ref5,pap1,ref6}.

The ratio of our result ${\mathcal B}^*_{D^+\to K^+\pi^+\pi^-\pi^0}$
over the world averaged value of ${\mathcal B}_{D^+\to K^-\pi^+\pi^+\pi^0}$
is $(1.81\pm0.15)\%$, corresponding to $(6.28\pm0.52)\tan^4\theta_C$,
where $\sin\theta_C = 0.2257$ \cite{pdg2020}.
This ratio is significantly larger than the values
(0.21-0.58)\% measured for the other DCS decays,
$D^0\to K^+\pi^-$, $D^0\to K^+\pi^-\pi^-\pi^+$,
$D^0\to K^+\pi^-\pi^0$, $D^+\to K^+\pi^+\pi^-$,
$D_s^+\to K^+K^+\pi^-$, and $\Lambda_c^+\to p  K^+\pi^-$~\cite{pdg2020}.
It is already known that the ratio
of ${\mathcal B}_{D^0\to K^+\pi^-\pi^-\pi^+}/{\mathcal B}_{D^0\to K^-\pi^+\pi^+\pi^-}$
roughly supports $\tan^4\theta_C$ \cite{pdg2020}.
This unexpected ratio implies that there is a massive isospin symmetry violation in the decays $D^+\to K^+\pi^+\pi^-\pi^0$ and $D^0\to K^+\pi^-\pi^-\pi^+$,
which may be caused by final state interactions and very different resonance structures in these two decays. Amplitude analyses of these decays with larger data samples~\cite{whitepaper}
will provide crucial information to understand the origin of the anomalously large ratio.
The asymmetry of the BFs of charge-conjugated decays $D^\pm \to K^\pm\pi^\pm\pi^\mp\pi^0$ is determined, and no evidence for $CP$ violation is found.

The authors thank Prof. Fu-Sheng Yu for helpful discussions.
The BESIII collaboration thanks the staff of BEPCII and the IHEP computing center for their strong support. This work is supported in part by National Key Basic Research Program of China under Contract No. 2015CB856700; National Natural Science Foundation of China (NSFC) under Contracts Nos. 11805037, 11775230, 11625523, 11635010, 11735014, 11822506, 11835012, 11935015, 11935016, 11935018, 11961141012; the Chinese Academy of Sciences (CAS) Large-Scale Scientific Facility Program; Joint Large-Scale Scientific Facility Funds of the NSFC and CAS under Contracts Nos. U1832121, U1732263, U1832207; CAS Key Research Program of Frontier Sciences under Contracts Nos. QYZDJ-SSW-SLH003, QYZDJ-SSW-SLH040; 100 Talents Program of CAS; INPAC and Shanghai Key Laboratory for Particle Physics and Cosmology; ERC under Contract No. 758462; German Research Foundation DFG under Contracts Nos. Collaborative Research Center CRC 1044, FOR 2359, FOR 2359, GRK 214; Istituto Nazionale di Fisica Nucleare, Italy; Ministry of Development of Turkey under Contract No. DPT2006K-120470; National Science and Technology fund; Olle Engkvist Foundation under Contract No. 200-0605; STFC (United Kingdom); The Knut and Alice Wallenberg Foundation (Sweden) under Contract No. 2016.0157; The Royal Society, UK under Contracts Nos. DH140054, DH160214; The Swedish Research Council; U. S. Department of Energy under Contracts Nos. DE-FG02-05ER41374, DE-SC-0012069.

\end{document}

%% file: BESIII_authors.tex
\author{
M.~Ablikim$^{1}$, M.~N.~Achasov$^{10,c}$, P.~Adlarson$^{64}$, S. ~Ahmed$^{15}$, M.~Albrecht$^{4}$, A.~Amoroso$^{63A,63C}$, Q.~An$^{60,48}$, ~Anita$^{21}$, X.~H.~Bai$^{54}$, Y.~Bai$^{47}$
% ...
\\
\vspace{0.2cm}
(BESIII Collaboration)\\
\vspace{0.2cm} {\it
$^{1}$ Institute of High Energy Physics, Beijing 100049, People's Republic of China\\
% ...
$^{4}$ Bochum Ruhr-University, D-44780 Bochum, Germany\\
% ...
$^{10}$ G.I. Budker Institute of Nuclear Physics SB RAS (BINP), Novosibirsk 630090, Russia\\
% ...
$^{15}$ Helmholtz Institute Mainz, Johann-Joachim-Becher-Weg 45, D-55099 Mainz, Germany\\
% ...
$^{21}$ Indian Institute of Technology Madras, Chennai 600036, India\\
% ...
$^{47}$ Southeast University, Nanjing 211100, People's Republic of China\\
$^{48}$ State Key Laboratory of Particle Detection and Electronics, Beijing 100049, Hefei 230026, People's Republic of China\\
% ...
$^{54}$ University of Jinan, Jinan 250022, People's Republic of China\\
% ...
$^{60}$ University of Science and Technology of China, Hefei 230026, People's Republic of China\\
% ...
$^{63}$ (A)University of Turin, I-10125, Turin, Italy; (B)University of Eastern Piedmont, I-15121, Alessandria, Italy; (C)INFN, I-10125, Turin, Italy\\
$^{64}$ Uppsala University, Box 516, SE-75120 Uppsala, Sweden\\
% ...
$^{c}$ Also at the Novosibirsk State University, Novosibirsk, 630090, Russia\\
% ...
}
}

%% file: submit_v2.bbl
\begin{thebibliography}{**}
\bibitem{note_BF}
This contribution is calculated by
$\sum\limits_{{\mathcal R}=1}^{3}\left [{\mathcal B}({D^+\to K^+{\mathcal R}})\cdot {\mathcal B}({\mathcal R}\to \pi^+\pi^-\pi^0)\right ]$,
where ${\mathcal R}$ sums over $\eta$, $\omega$, and $\phi$,
${\mathcal B}(D^+\to K^+\omega)$ is obtained in this work, and the other BFs are quoted from the PDG~\cite{pdg2020}.
\end{thebibliography}
